\documentclass[11pt,letterpaper]{article}

\usepackage{typearea}
\typearea{15}

\usepackage{theorem,latexsym,graphicx,amssymb}
\usepackage{amsmath,enumerate}
\usepackage{wrapfig}
\usepackage{subfigure}
\usepackage{psfig}
\usepackage{epsfig}
\usepackage{xspace}
\usepackage{paralist}
\usepackage[compact]{titlesec}
\usepackage[boxed,lined]{algorithm2e}

\usepackage{color}
\definecolor{Darkblue}{rgb}{0,0,0.4}
\definecolor{Brown}{cmyk}{0,0.81,1.,0.60}
\definecolor{Purple}{cmyk}{0.45,0.86,0,0}
\usepackage[breaklinks]{hyperref}
\hypersetup{colorlinks=true,
            citebordercolor={.6 .6 .6},linkbordercolor={.6 .6 .6},%
citecolor=black,urlcolor=black,linkcolor=black,pagecolor=black}
\newcommand{\lref}[2][]{\hyperref[#2]{#1~\ref*{#2}}}


\usepackage{multicol}

\newenvironment{proof}{{\bf Proof:  }}{\hfill\rule{2mm}{2mm}}

\newtheorem{theorem}{Theorem}[section]
\newtheorem{lemma}{Lemma}[section]

\newtheorem{fact}{Fact}[section]

\newtheorem{proposition}{\hskip\parindent Proposition}[section]

\providecommand{\Appendix}{}
\renewcommand{\Appendix}[2][?]{%
        \refstepcounter{section}%
        \vspace{\parskip}%
        {\flushright\large\bfseries\appendixname\ \thesection: #1}%
        \vspace{\baselineskip}%
}

\renewcommand{\appendix}{%
        \newpage
        \renewcommand{\section}{\secdef\Appendix\Appendix}%
        \renewcommand{\thesection}{\Alph{section}}%
        \setcounter{section}{0}%
}

\newcommand{\eps}{\epsilon}

\newcommand{\R}{\mathbb{R}}

\newcommand{\ceil}[1]{\ensuremath{\lceil #1 \rceil}}

\newcounter{note}[section]
\renewcommand{\thenote}{\thesection.\arabic{note}}

\newcommand{\initOneLiners}{%
    \setlength{\itemsep}{0pt}
    \setlength{\parsep }{0pt}
    \setlength{\topsep }{0pt}
}

\newcommand{\ignore}[1]{}

\newcommand{\shortv}[1]{}

\usepackage[compact]{titlesec}

\newcommand{\floor}[1]{\ensuremath{\left\lfloor#1\right\rfloor}}
\renewcommand{\dim}{\textnormal{dim}}

\newcommand{\MST}{\mathsf{MST}}

\newcommand{\length}{\mathsf{length}}

\newcommand{\Add}{\mathsf{Add}}

\providecommand{\Appendix}{}
\renewcommand{\Appendix}[2][?]{%
        \refstepcounter{section}%
        \vspace{\parskip}%
        {\flushright\large\bfseries\appendixname\ \thesection: #1}%
        \vspace{\baselineskip}%
}

\renewcommand{\appendix}{%
        \newpage
        \renewcommand{\section}{\secdef\Appendix\Appendix}%
        \renewcommand{\thesection}{\Alph{section}}%
        \setcounter{section}{0}%
}

\newcommand{\mingfei}[1]{\refstepcounter{note}$\ll${\sf Mingfei's
Comment~\thenote:} {\sf \textcolor{green}{#1}}$\gg$\marginpar{\tiny\bf MC~\thenote}}

\newcommand{\hubert}[1]{\refstepcounter{note}$\ll${\sf Hubert's
Comment~\thenote:} {\sf \textcolor{red}{#1}}$\gg$\marginpar{\tiny\bf HC~\thenote}}

\newcommand{\li}[1]{\refstepcounter{note}$\ll${\sf Li's
Comment~\thenote:} {\sf \textcolor{blue}{#1}}$\gg$\marginpar{\tiny\bf LC~\thenote}}

\title{Incubators vs Zombies: Fault-Tolerant, Short, Thin and Lanky Spanners for Doubling Metrics}

\author{T-H. Hubert Chan \and Mingfei Li \and Li Ning
\thanks{Department of Computer Science, The University of Hong Kong}
\thanks{\{hubert, mfli, lning\}@ cs.hku.hk}
}
\date{}

\begin{document}

\begin{titlepage}

\maketitle

\begin{abstract}

Recently Elkin and Solomon gave a construction
of spanners for doubling metrics that has constant
maximum degree, hop-diameter $O(\log n)$ and lightness $O(\log n)$ (i.e.,
weight $O(\log n) \cdot w(\textsf{MST}))$.
This resolves a long standing conjecture
proposed by Arya et al. in a seminal STOC 1995 paper.

However, Elkin and Solomon's spanner construction is
extremely complicated; we offer a simple alternative construction
that is very intuitive and is based on the standard
technique of net tree with cross edges.  Indeed, our approach
can be readily applied to our previous construction of $k$-fault tolerant
spanners (ICALP 2012) to achieve $k$-fault tolerance,  maximum
degree $O(k^2)$, hop-diameter $O(\log n)$ and lightness $O(k^3 \log n)$.

\end{abstract}

\thispagestyle{empty}
\end{titlepage}

\section{Introduction}

A finite metric space $(X, d)$ with $n = |X|$ can be represented by a complete graph $G = (X, E)$, where the edge weight $w(e)$ on an edge $e = \{x, y\}$ is $d(x, y)$.
A \emph{$t$-spanner} of $X$, is a weighted subgraph $H = (X, E')$ of $G$ that preserves all pairwise distance within a factor of $t$,
i.e., $d_H(x, y) \leq t \cdot d(x, y)$ for all $x, y \in X$,
where $d_H(x, y)$ denotes the shortest-path distance between $x$ and $y$ in $H$,
and the factor $t$ is called the $stretch$ of $H$.
A path between $x$ and $y$ in $H$ with length at most $t \cdot d(x, y)$ is called a \emph{$t$-spanner path}.
Spanners have been studied extensively since the mid-eighties
(see \cite{DBLP:conf/soda/CallahanK93,DBLP:conf/compgeom/DasN94,Arya1995,DBLP:conf/esa/SolomonE10,DBLP:conf/compgeom/Har-PeledM05,DBLP:conf/soda/ChanGMZ05,DBLP:journals/dcg/ChanG09,DBLP:conf/esa/GottliebR08} and the references therein; also refer to
\cite{DBLP:books/daglib/0017763} for an excellent survey), and find applications in approximation algorithms, network topology design, distance oracles, distributed systems.

Spanners are important structures, as they enable approximation of a metric space in a much more economical form.
Depending on the application, there are parameters of the spanner other than stretch
that can be optimized.  The total weight of the edges should be
at most some factor (known as \emph{lightness}) times the weight
of a minimum spanning tree (MST) of the metric space.  It might
also be desirable for the spanner to have small maximum degree (hence also
having small number of edges), or small 
\emph{hop-diameter},
i.e., every pair of points $x$ and $y$ should be connected by a $t$-spanner path with a small number of edges.

Observe that for some metric spaces such as the uniform metric on $n$ points, the only possible spanner with stretch 1.5 is the complete graph. \emph{Doubling metrics} are special classes of metrics, but still have interesting properties. The \emph{doubling dimension} of a metric space $(X, d)$, denoted by $\dim(X)$ (or $\dim$ when the context is clear),
is the smallest value $\rho$ such that every ball in $X$ can be covered by $2^\rho$ balls of half the radius \cite{DBLP:conf/focs/GuptaKL03}.
A metric space is called \emph{doubling}, if its doubling dimension is bounded by some constant.
Doubling dimension is a generalization of Euclidean dimension to arbitrary metric spaces,
as the space $\R^T$ equipped with $\ell_p$-norm  has doubling dimension $\Theta(T)$ \cite{DBLP:conf/focs/GuptaKL03}.
Spanners for doubling metrics have been studied
in \cite{DBLP:conf/compgeom/Har-PeledM05,DBLP:conf/soda/ChanGMZ05,DBLP:journals/dcg/ChanG09,DBLP:conf/esa/GottliebR08,DBLP:conf/esa/SolomonE10}.

Sometimes we want our spanner to be robust against node failures, meaning that even when some of the nodes in the spanner fail,
the remaining part is still a $t$-spanner.
Formally, given $1 \leq k \leq n - 2$, a spanner $H$ of $X$ is called a $k$-vertex-fault-tolerant $t$-spanner
($(k, t)$-VFTS or simply $k$-VFTS if the stretch $t$ is clear from context),
if for any subset $S \subseteq X$ with $|S| \leq k$, $H \setminus S$ is a $t$-spanner for $X \setminus S$.

\noindent \textbf{Our Contributions.}  Our main theorem subsumes the results of
two recent works on spanners for doubling metrics: (1) our previous paper~\cite{Chan2012} on fault-tolerant spanners
with constant maximum degree or small hop-diameter, 
(2) Elkin and Solomon's spanner construction~\cite{SolomonFOCS12} with constant maximum degree, $O(\log n)$ hop-diameter
and $O(\log n)$ lightness.

\begin{theorem}[VFTS with $O(1)$ Max Degree, $O(\log n)$ Hop-Diameter, $O(\log n)$ Lightness]
\label{thm:main}
Let $(X, d)$ be a metric space with $n$ points, and let $0 < \eps < 1$ be a constant.
Given $1 \leq k \leq n - 2$, there exists a $(k, 1+ \eps)$-VFTS with 
maximum degree $\eps^{-O(\dim)}\cdot k^2$,
hop-diameter $O(\log n)$,
and lightness $\eps^{-O(\dim)}\cdot k^3 \cdot \log n$.
\end{theorem}

\noindent \emph{Research Background.}
We review the most relevant related work; the readers can refer to~\cite{Chan2012,SolomonFOCS12}
for more references.
In a seminal STOC 1995 paper, Arya et al.~\cite{Arya1995} gave several constructions for \emph{Euclidean spanners} 
that trade between the number of edges, maximum degree,
hop-diameter and lightness.
In particular, they showed that for any $n$ points in low-dimensional Euclidean space,
there exists a $(1 + \eps)$-spanner with $O(n)$ edges,
constant maximum degree, hop-diameter $O(\log n)$ and lightness $O(\log^2 n)$.
Since then, it has been a long standing open problem whether there exists a Euclidean $(1 + \eps)$-spanner
with lightness $O(\log n)$ and all other properties above; this is
the best possible lightness because of a lower bound $\Omega(\log n)$ result by Dinitz et al.~\cite{DBLP:conf/focs/DinitzES08}.

Seventeen years later, Elkin and Solomon~\cite{SolomonFOCS12} recently answered this open problem in the affirmative,
and showed a stronger result: their construction actually works for doubling metrics.
Perhaps it might not come as a surprise that their construction is highly complicated.  The construction
starts with an Euler
tour on an MST and partitions the tour into hierarchical blocks, with very sophisticated
nomenclature systems to describe various kinds of counters and representatives for the blocks, in order to
reassign parent-child relationship between the blocks for achieving a spanner with all the
desirable properties.

While we were trying to study the intricate rules of their construction\footnote{Even to this date,
we do not claim that we fully understand the construction in~\cite{SolomonFOCS12}.},
we began to look for a simpler construction.  We discovered that
the standard net tree with cross edges framework (used in~\cite{DBLP:conf/soda/ChanGMZ05})
can be augmented with a few modifications to give a spanner construction with the same desirable properties.  Since our
fault-tolerant spanners~\cite{Chan2012} are also based on
net trees, our techniques can be used
for constructing fault-tolerant spanners with similar maximum degree,
hop-diameter and lightness guarantees as well.

We believe that it is an important contribution to have
a simple and intuitive construction, which can be readily
understood and used by researchers in the community. For instance,
it is not immediately clear how Elkin and Solomon's construction can be
used to construct fault-tolerant spanners.  As a way of
paying tribute to their work~\cite{SolomonFOCS12}, we borrow their
terminology \emph{incubator} and \emph{zombie} (some of
our zombies can also \emph{disappear}!), although their
meanings are totally different in this paper.

\noindent \textbf{Our Techniques.} In order
to illustrate our techniques, we now give the main ideas
for constructing a spanner when all nodes are functioning.  The formal
treatment for fault-tolerance is given in the main body of the paper.
We use the standard net tree with cross edges framework given
in~\cite{DBLP:conf/soda/ChanGMZ05}: given a metric
space $(X,d)$, construct a hierarchical
sequence $\{N_i\}$ of nets with geometrically decreasing distance scales, and
build a tree structure, where the root has only 1 point with distance scale
around the maximum inter-point distance $\Delta$, and each leaf corresponds
to each point in $X$ at distance scale around the minimum inter-point distance (assumed
to be more than 1).
At each level, cross edges are added between net points that are
close to each other relative to their level distance scale.  A basic spanner~\cite{DBLP:conf/soda/ChanGMZ05}
consisting of the tree edges and the cross edges can be shown to have low stretch.  The basic idea is that for any two points
$u$ and $v$, we can start at the corresponding leaf nodes and climb
to an appropriate level depending on $d(x,y)$ to net points $u'$ and $v'$ that
are close to $u$ and $v$ respectively such that the cross edge $\{u',v'\}$ is guaranteed to exist.  We analyze each
of the properties: maximum degree, hop-diameter and lightness, and
describe how issues arise and can be resolved.

\noindent \emph{Lightness.}  For doubling metrics, lightness comes almost for free (see details in Section~\ref{sec:diam}).  Since the sum of all edges with length at most $\frac{\Delta}{n^2}$
is at most $\Delta$, which is at most the weight of an MST, the significant distance
scales are those between $\frac{\Delta}{n^2}$ and $\Delta$, and
there are $O(\log n)$ of them.  The standard analysis in~\cite{DBLP:conf/soda/ChanGMZ05} uses doubling dimension to argue
that at each level, each net point only has constant number of neighbors at that level.  Hence, one can conclude that the weight of edges from each significant
distance scale is a constant times that of an MST, thereby giving $O(\log n)$ lightness.

\noindent \emph{Maximum Degree.} Using doubling dimension,
it is shown in~\cite{DBLP:conf/soda/ChanGMZ05} that edges
can be directed such that the out-degree of every node is constant.
For each node $v$, the edges from its incoming neighbors
are replaced by a constant degree single-sink spanner rooted at $v$.  This idea
is used in~\cite{DBLP:conf/soda/ChanGMZ05} and made explicit in~\cite{Chan2012}.
By the low stretch property of the single-sink spanner,
the weight will increase by only a constant factor.
Although no non-trivial bounds on the hop-diameter of 
the single-sink spanners are given
in~\cite{DBLP:conf/soda/ChanGMZ05,Chan2012}, an easy modification
is described in Section~\ref{sec:degree} to achieve $O(\log n)$ hop-diameter.
Since the out-degree of every node is constant, it can only participate
in only a constant number of single-sink spanners.  Therefore, after this 
single-sink transformation for every node, the maximum degree of the resulting
spanner is constant.

\noindent \emph{Hop-Diameter.}  If the metric space has
 large maximum distance $\Delta$ with respect to the minimum inter-point distance,
 the number of levels in the net tree can be large.  In particular,
 in the above mentioned spanner path between $u$ and $v$, it can take
 many hops to go from $u$ to $u'$ in the appropriate level.  However, this can be easily fixed by adding shortcutting edges to subtrees at  distance scales less than $\frac{\Delta}{n^2}$ via the 1-spanner construction
 for tree metrics by Solomon and Elkin~\cite{DBLP:conf/esa/SolomonE10}
 with $O(\log n)$ hop-diameter and $O(\log n)$ lightness; moreover,
 this 1-spanner construction only increases the maximum degree of the input tree by a constant.  Since the edges from insignificant distance scales
 have total weight at most that of an MST, this will still give $O(\log n)$ lightness.
 
 There are only $O(\log n)$ significant distance scales and the
 tree edges from insignificant distance scales can be climbed using $O(\log n)$ hops.  Hence, it might seem that
 we have $O(\log n)$ hop-diameter.  However, observe that each tree edge
 might participate in a single-sink spanner and might be replaced by
 a path with $\Omega(\log n)$ hops.  Therefore, we can only bound the
 hop-diameter by $O(\log^2 n)$.  
 
 If we could stop tree edges from participating
 in single-sink spanners, then the hop-diameter would be $O(\log n)$.  However,
 the issue is that a net tree is likely to have large degree, because
 of the hierarchical property of the nets.  In particular, the root node is a net point at every level, and hence could have many tree edges connecting to it.  Our
 key idea is simple: we keep the net tree structure, but replace each node with another representative (that is close by) such that the degree due to tree edges induced on the representatives is constant.  To describe our ideas clearly, we use the terminology \emph{incubator}
 for a node in the net tree and \emph{zombie} for a representative.

\noindent \emph{Incubators Working with Zombies.} If a point $x$ in $X$ is a
net point in several levels, we explicitly have a separate node known as
an \emph{incubator} for each level $x$ is in. The net tree structure before
becomes a tree on the incubators, and a former cross edge at some level
becomes an incubator edge at the same level.
  If an internal incubator has only one child, we merge the two incubators together
  and the merged incubator inherits the corresponding edges; this merging is repeated until every internal incubator has at least two children.  Observe that each incubator
  only has a constant number of tree edges incident to it.  
  
Each incubator has a representative known as a \emph{zombie}, each of which
is identified with a point in $X$.  The graph structure on incubators naturally
induces a spanner on $X$: if there is an edge between two incubators, then there
is an induced edge between the corresponding zombies.  The remaining issue is how we
assign zombies to incubators; the following procedure turns out to be sufficient.

\noindent \emph{``The zombies are climbing...''}  Each leaf incubator initially
has two zombies, both having the same identity as the leaf.  One zombie stays at the leaf and the other zombie climbs up the tree.  The order of zombie climbing is arbitrary.  If a zombie finds an empty incubator, it stays there, and otherwise continues to climb up; if the root incubator is reached and is already occupied, the zombie simply \emph{disappears}.

\noindent \emph{Sketch Analysis.}  After merging, each internal incubator
has at least two children.  Hence, there will be enough zombies such that
each incubator will be occupied by a zombie that originates from one of its descendant leaves. We are essentially replacing each net point corresponding
to an incubator with a point represented by the zombie that is
close by with respect to the relevant distance scale.  It can be shown
that the stretch is still preserved.  Moreover, observe that every point in $X$
can only be the identity of at most two zombies.  Since the incubator tree
has constant degree, it follows that the degree of a point due to induced tree
edges is constant.  Hence, we can let only the edges not induced by tree edges
to participate in the single-sink spanner construction.  This resolves
the aforementioned issue of the net tree having high degree, and so $O(\log n)$ hop-diameter can be achieved.

We apply the incubator-zombie technique to the fault-tolerant spanner construction
in~\cite{Chan2012}.  The rules for inducing edges on $X$ are slightly more
involved and are given in Section~\ref{sec:zombie}.

\noindent \textbf{Future Direction.}  Observe that
for our $k$-fault tolerant spanner construction,
the dependence on $k$ for maximum degree is $O(k^2)$
and for lightness is $O(k^3)$, where the lower bounds
are $\Omega(k)$ (trivial) and $\Omega(k^2)$ (a simple example
is given in~\cite{DBLP:journals/dcg/CzumajZ04}) respectively;
it is interesting to see whether the dependence on $k$ can be improved.
Indeed, for low-dimensional Euclidean metrics, Czumaj and Zhou~\cite{DBLP:journals/dcg/CzumajZ04} constructed $k$-fault tolerant
spanners with maximum degree $O(k)$ and $O(k^2)$ lightness, but
no guarantee on hop-diameter.  For doubling metrics,
another interesting open question is to construct spanners with constant lightness 
even when no nodes fail and with no restriction on degree or hop-diameter.

\section{Preliminaries}
\label{sec:prelim}

For any positive integer $m$, we denote $[m] := \{1, 2, \ldots, m\}$.
Throughout this paper, let $(X, d)$ be a doubling metric with $n$ points, $1 \leq k \leq n - 2$ be an integer representing the number of failed nodes allowed,
and let $0 < \eps < \frac{1}{2}$ be a constant and consider $1+ \eps$ stretch.
Without loss of generality, we also assume that the minimum inter-point distance of $X$ is greater than $1$.
We denote $\Delta := \max_{x, y \in X}d(x, y)$ as the \emph{diameter} of $X$.

Suppose $r > 0$. The ball of radius $r$ centered at $x$ is $B(x, r) := \{y \in X : d(x, y) \leq r\}$.
We say that a cluster $C \subseteq X$ has radius at most $r$, if there exists $x \in C$ such that $C \subseteq B(x, r)$.
Let $r_2 > r_1 > 0$. 
The ring of inner radius $r_1$ and outer radius $r_2$ centered at $x$ is $R(x, r_1, r_2) := B(x, r_2) \setminus B(x, r_1)$. 



A set $Y \subseteq X$ is an \emph{$r$-cover} of $X$ if for any point $x \in X$ there is a point $y \in Y$ such that $d(x, y) \le r$.
A  set $Y$ is an \emph{$r$-packing} if for any pair of distinct points $y, y' \in Y$, it holds that $d(y, y') > r$. 
We say that a set $Y \subseteq X$ is an \emph{$r$-net} for $X$ if $Y$ is both an $r$-cover of $X$ and an $r$-packing. 
Note that if $X$ is finite, an $r$-net can be constructed by a greedy algorithm. 
By recursively applying the definition of doubling dimension, we can get the following key proposition \cite{DBLP:conf/focs/GuptaKL03}.

\begin{proposition}[Nets Have Small Size~\cite{DBLP:conf/focs/GuptaKL03}] \label{prop:small_net}
Let $R \geq 2r > 0$ and let $Y \subseteq X$ be an $r$-packing contained in a ball of radius $R$. Then, $|Y| \le (\frac{R}{r})^{2\dim}$.
\end{proposition}

The minimum spanning tree of a finite metric space $(X,d)$ is denoted by $\MST(X)$ 
(or simply $\MST$ if $(X,d)$ is clear).
Also, we use $w(\MST)$ to denote the weight of $\MST$.
Given a spanner $H$ on $(X,d)$, the \emph{lightness} of $H$ is defined
by the ratio of the total weight of edges in $H$ to the weight of $\MST$.

\begin{fact}[Lower Bounds for $\MST$]\label{fact}
We have the following two lower bounds for $w(\MST)$.
\begin{compactitem}
\item[1.]
The weight of $\MST$ is bounded from below by its diameter $\Delta$.
\item[2.]
Let $S \subseteq X$ be an $r$-packing, with $r\leq \Delta$. Then,  $w(\MST) \geq \frac{1}{2}r\cdot|S|$.
\end{compactitem}
\end{fact}

The fault-tolerant spanner construction in \cite{Chan2012},
which is the basis for our construction in this paper,
relies on \emph{fault-tolerant hierarchical nets}. We review their construction and some key properties of them.

\noindent \textbf{Fault-Tolerant Hierarchical Nets.}
We color each point in $X$ with one of the colors in $[k + 1]$, and let $X_c$ be the set of points with color $c$.
For each color $c \in [k + 1]$, we build a sequence of hierarchical nets of $\ell := \ceil{\log_2 \Delta}$ levels,
$X_c = N^c_0 \supseteq N^c_1 \supseteq \cdots \supseteq N^c_\ell$.
We denote by $N_i := \cup_{c \in [k + 1]}N_i^c$ the set of all level-$i$ net points. Let $r_i := 2^i$ be the \emph{distance scale} of level~$i$.
Fault-tolerant hierarchical nets should satisfy the following properties:
\begin{compactitem}
\item[1.] \emph{Packing.} For each $0 \leq i \leq \ell$ and $c \in [k + 1]$, $N_i^c$ is an $r_i$-packing;
\item[2.] \emph{Covering.} For any $1 \leq i \leq \ell $, if $x \in X$ is not a net point in $N_i$,
then for each color $c \in [k + 1]$, there exists a net point $y_c\in N^c_i$ such that $d(y_c, x) \leq r_i$.
\end{compactitem}

\noindent \emph{Construction.} The hierarchical nets can be constructed in a top-down approach. Initially, each $N^c_\ell$ consists of a distinct point in $X$.
Note that $k \leq n - 2$ and hence the initialization is well defined.
Also, the single point in $N^c_{\ell}$ is colored with $c$ and points not included in any cluster $N^c_\ell$ stay uncolored.

Suppose all nets on level $i + 1$ have been built and we construct the level-$i$ nets as follows.
For $c$ from $1$ to $k + 1$, let $U_c$ be the set of uncolored points when we start to build $N_i^c$, i.e., after finishing the construction of $N_i^1, N_i^2, \ldots, N_i^{c-1}$.
We initialize $N_i^c := N_{i+1}^c$, and extend $N_{i+1}^c$ to get $N_i^c$ by
greedily adding points in $U_c$ to $N_i^c$ such that the resulting $N_i^c$ is an $r_i$-net for $U_c$; we color the points in $N_i^c \cap U_c$ with color $c$.

Note that the packing property and the covering property follow directly from the net construction.

\section{Incubators Working with Zombies: Reducing Degree of Net Trees}
\label{sec:zombie}

As mentioned in the introduction, the net-tree-with-cross-edges approach in~\cite{DBLP:conf/soda/ChanGMZ05} has the
issue that the degree of a net tree can be high.
The reason is that a high level net point (in particular the root) can participate
in many levels in the net tree, causing high degree.
The use of single-sink spanner can reduce the maximum degree, but at the cost of replacing
a tree edge with a path consisting of $\Omega(\log n)$ hops.
In this section, we modify the net tree construction such that
each point participates in a small number of levels.  We borrow
some terminology from~\cite{SolomonFOCS12} and define our own
\emph{incubators} and \emph{zombies} to describe our construction,
which is based on the fault-tolerant hierarchical nets defined in
Section~\ref{sec:prelim}.

\ignore{
In this section, we introduce the concept of \emph{incubator trees}, which are a variant of the standard net trees.
Combining with \emph{climbing zombies},
we give a construction of $(1 + \eps)$-spanner induced by the incubator trees and cross edges such that
the degree due to tree edges is bounded.
}

\noindent \textbf{Incubators.}  For each level $i$, and each $x \in N_i$,
there is a corresponding \emph{incubator} $y=(x,i)$, where $x$ is the \emph{identity} of the incubator
and $i$ is its level.  The color of an incubator is the same as its identity.
Observe that the original metric induces a distance function
on the incubators in a natural way: if $y_1 = (x_1, i_1)$ and $y_2 = (x_2, i_2)$, then $d(y_1, y_2) = d(x_1, x_2)$.
We also define distances between points in the original metric space and incubators:
if $y_1 = (x_1, i_1)$ and $x_2 \in X$, then $d(y_1, x_2) = d(x_1, x_2)$.

\noindent \textbf{Incubator Graph.}  We define a graph $\mathcal{H}$ on incubators.  An edge is \emph{local}
if it connects two incubators of the same color, and is \emph{foreign} otherwise.  Edges are added between
incubators according to the following rules.

\begin{compactitem}
\item[1.] \emph{Tree edges.}  For each level $0 \leq i < l$, color $c \in [k+1]$, and $x \in N_i^c$, consider
the incubator $y = (x,i)$. Suppose $x_c \in N_{i+1}^c$ is a closest point in $N_{i+1}^c$ to $x$. (Observe that
if $x \in N_{i+1}^c$, then $x = x_c$; if $x \notin N_{i+1}^c$, then $d(x, x_c) \leq r_{i+1}$.)
Then, we add a local tree edge between the incubators $y$ and $y_c = (x_c, i+1)$, and
we call $y_c$ the \emph{local  parent} of $y$ and $y$ is a \emph{local child} of $y_c$.   All the incubators
of color $c$ form an \emph{incubator tree} $T_c$.

Suppose $x \in N_i^c \setminus N_{i+1}^c$.  Then, for each color $c' \neq c$,
suppose $x_{c'} \in N_{i+1}^{c'}$ is a closest point in $N_{i+1}^{c'}$ to $x$. 
(Note that $d(x, x_{c'}) \leq r_{i+1}$.)
Then, we add a foreign tree edge between the incubators $y=(x,i)$ and $y_{c'} = (x_{c'}, i+1)$.
Then, $y_{c'}$ is a \emph{foreign parent} of $y$ and $y$ is a \emph{foreign child} of $y_{c'}$.

\item[2.] \emph{Cross edges.} For each level $0 \leq i \leq l$, and all $u \neq v \in N_i$
such that $d(u,v) \leq \gamma \cdot r_i$, where $\gamma := 34 + \frac{272}{\eps}$
as in~\cite{DBLP:conf/soda/ChanGMZ05,Chan2012}, we add
a cross edge between the incubators $(u,i)$ and $(v,i)$. 

\item[3.] \emph{Skeleton edges.}  The idea in~\cite{DBLP:conf/soda/ChanGMZ05,Chan2012}
to obtain a low-stretch path between two points is to use a similar net tree structure: for both points,
start at their corresponding leaves and climb up the tree to a certain level,
at which level the net points are connected by a cross edge.  However,
if the number of levels is large, in particular when the diameter of the metric space
is super-polynomial, then the number of hops in the path can be large.  To resolve
this issue, we use techniques in~\cite{DBLP:conf/esa/SolomonE10} to shortcut
each incubator tree $T_c$ by adding more local edges; the details are given
in Section~\ref{sec:diam}.  \emph{Skeleton edges} are the local tree
edges together with the local shortcutting edges that we add later;
in this section, we only have local tree edges as skeleton edges.
\end{compactitem}

\noindent \textbf{Lonely vs Super Incubators.} An incubator is \emph{lonely} if it has exactly one local child incubator (which has the same identity as the parent), otherwise
it is \emph{normal}.  Observe that the leaf incubators have no children and are normal.  Whenever
there is a lonely incubator, we \emph{merge} it with its only local child to form a \emph{super incubator},
which inherits the edges from the two merged incubators.  This process continues if the resulting
super incubator is still lonely; hence, we combine a chain of lonely incubators (with a normal incubator at
the bottom of the chain) to form a super incubator.  A super incubator is described
by $y = (x, I)$, where $I$ is the set of levels from which incubators are merged to form $y$.
Observe
that after incubator \emph{merging}, there are only normal and super incubators, and every
non-leaf incubator has at least two local child incubators.

For the rest of this section,
 the incubator graph $\mathcal{H}$ is the union of tree edges and cross edges after incubator merging;
 we later augment this graph by adding local shortcutting edges in Section~\ref{sec:diam}.
Moreover, we use incubator trees to refer to the trees after incubator merging.
Also, 
when we refer to an incubator $(x, i)$, if it has been merged into a super incubator,
we are actually referring to the corresponding super incubator.

Note that the incubator graph $\mathcal{H}$ induces a spanner on $X$ naturally:
an edge between incubators $y_1 = (x_1, i_1)$ and $y_2 = (x_2, i_2)$ induces an edge 
between $x_1$ and $x_2$.  The spanner recovered by the above incubator description
is essentially the one constructed in \cite{Chan2012}.  As mentioned earlier,
we would like to reduce the degree due to local tree edges, and we
use the concept \emph{zombies} to assign representatives to incubators.


\ignore{
\noindent\textbf{Incubator Edges.} For two incubators $y_1=(x_1, i_1)$ and $y_2=(x_2, i_2)$,
it induces an \emph{incubator edge} $\{x_1,x_2\}$ between $x_1$ and $x_2$, if $\{y_1,y_2\}$ is an edge in some
net tree with incubators, or an cross edge. Hence the construction of net trees with incubators and cross edges
induce a spanner structure over points in $X$.
}

\noindent \textbf{Zombies.} 
A \emph{zombie} is identified by a point $x$ in the metric space,
which is the \emph{identity} of the zombie.
The color of a zombie is the same as its identity.
When the context is clear,
we do not distinguish between a zombie and its identity. Each incubator
is occupied by exactly one zombie with the same color
 in the following procedure.

\noindent \textbf{``The zombies are climbing...''}
%
Initially, each leaf incubator $y = (x, 0)$ at level 0
contains a zombie with the same identity $x$; the incubator clones
an extra copy of the zombie, which climbs up the incubator tree $T_c$
of its color in the following way.  The order in which zombies climb is arbitrary.
If a zombie enters an incubator which contains no zombie, then the zombie
occupies the incubator; we say that a zombie is at level $i$ if it occupies
an incubator at level $i$. In particular, if a zombie occupies a super incubator
$y = (x,I)$, we say the zombie is at all levels with indices in $I$.
Otherwise, the current incubator already contains another
zombie (this applies to the case when a cloned zombie finds
another zombie already contained in the leaf incubator); in this case,
the zombie climbs to the local parent of the current incubator in the incubator
tree $T_c$.  If the current incubator is the root and is already occupied,
the zombie simply \emph{disappears}.  


Observe that in each incubator tree, each internal incubator
has at least two children, and hence there are enough zombies such that every incubator
will be occupied by one zombie with the same color.
Moreover, each incubator $y$ contains a zombie
which has the same identity as one of the local descendant leaf incubators of $y$.

\noindent \textbf{Spanner Induced by Incubator Graph with Zombies.}
The incubator graph $\mathcal{H}$ with zombies induces a spanner $H$ for the original space $X$ as follows.
Suppose there is an edge between incubators
 $y_1 = (x_1, i_1)$ and $y_2 = (x_2, i_2)$, which contain zombies $z_1$ and $z_2$ respectively.  If the edge
 $\{y_1, y_2\}$ is a skeleton edge (i.e. a local tree edge in this section),
 then an edge between the zombies $z_1$ and $z_2$ is induced; 
 otherwise (i.e. a foreign or a cross edge in this 
 section), 
 an edge between the identities $x_1$ and $x_2$ of the incubators is induced.
A path on $X$ induced by a path in the incubator tree $T_c$ is called a $c$-path.

\noindent \textbf{Naming Convention. (Please read!)}  We use the following
naming convention to describe induced edges on $X$.  Suppose
$\textsf{type}$ is a certain class for describing incubator edges, such as ``tree'', ``cross'',
``local'' or ``foreign''.  We say an edge on $X$ is a $\textsf{type}$
edge if it is induced by a $\textsf{type}$ edge between incubators.  For example,
an edge on $X$ is a cross edge if it is induced by a cross edge
between incubators.  Observe that an edge on $X$ can be induced
by multiple incubator edges.  We say an edge
on $X$ is a \emph{pure} $\textsf{type}$ edge if it is induced
by $\textsf{type}$ incubator edges \textbf{only}, and it is
a \emph{non}-$\textsf{type}$ edge if it is not induced
by any $\textsf{type}$ incubator edge.  For example,
if an edge on $X$ is induced by a cross edge on incubators,
but not by any tree edge, then it is both
a pure cross edge and a non-tree edge.



The properties of the induced graph $H$ are summarized in the following lemma, which is the main result of the section.
We first prove Property 2 and 3.

\begin{lemma}\label{lemma:basicH}
The induced graph
$H$ is a $(k, 1 + \eps)$-VFTS for $X$ with the following properties.
\begin{compactitem}
\item[1.] For any set $S \subseteq X$ with $|S| \leq k$, and any $x, y \in X \setminus S$,
there exists a $(1 + \eps)$-spanner path in $H \setminus S$ between $x$ and $y$ 
which is the concatenation of at most two foreign tree edges, 
at most one pure cross edge, and at most two $c$-paths for some color $c$ 
(though not necessarily in this order).
\item[2.] The maximum degree of $H$ due to local tree edges is bounded by $2^{O(\dim)}$.
\item[3.] The non-skeleton edges in $H$ can be directed such that the out-degree of 
any point in $H$ due to non-skeleton edges is bounded by $\eps^{-O(\dim)}\cdot k$.
\end{compactitem}
\end{lemma}


We first show that the maximum degree due to local tree edges is a constant.

\begin{lemma}[Small Degree due to Local Tree Edges]\label{lemma:local_degree}
For all $x \in X$, the degree of $x$ in $H$ due to local tree edges is $2^{O(\dim)}$.
\end{lemma}
\begin{proof}
We first show that each incubator has at most $4^{\dim}$ local children.
Let $y$ be any internal incubator of $T_c$.
Let $i$ be $y$'s level
(if $y$ is a super incubator, let $i$ be the lowest level in which $y$ participates).
Note that the identities of $y$'s children in $T_c$ form an $r_{i - 1}$-net,
and they are all within a distance of $r_i$ from $y$.
By Proposition~\ref{prop:small_net}, $y$ has at most $2^{2\dim}$ local children.

Note that for any $x \in X$,
there are two zombies identified by $x$,
each of which occupies at most one incubator.
Hence, the degree of $x$ due to local tree edges is bounded by $2(2^{2\dim} + 1) = 2^{O(\dim)}$.
\end{proof}

Now we show how to direct the non-skeleton edges in
the induced graph $H$; observe that a non-skeleton edge
is induced by either a foreign tree edge or a cross edge. If more than one rule applies,
we can pick the direction indicated by any one of the rules.

\begin{compactitem}
\item \emph{Foreign Tree Edges.} For a foreign tree edge
between a child $y_1 = (x_1,i_1)$ and a parent $y_2 = (x_2, i_2)$, we direct from $x_1$ to $x_2$.

\item \emph{Cross Edges.} Suppose $\{x_1, x_2\}$ is
a cross edge in $H$. 
Let $i^*(x)$ be the highest level $i$ such that $x \in N_i$.  The cross edge
$\{x_1, x_2\}$ is directed from $x_1$ to $x_2$ if $i^*(x_1) < i^*(x_2)$.
If $i^*(x_1) = i^*(x_2)$, the cross edge $\{x_1, x_2\}$ is directed arbitrarily.
\end{compactitem}

\begin{lemma}[Small Out-Degree due to Non-Skeleton Edges]\label{lemma:fc_degree}
If the non-skeleton edges in $H$ are directed as above, then
the out-degree due to them is bounded by $\eps^{-O(\dim)} \cdot k$.
\end{lemma}

\begin{proof}
Fix a point $x$ in $X$. we show that its out-degree due to foreign tree edges is at most $k$,
and its out-degree due to cross edges is at most $\eps^{-O(\dim)} \cdot k$.

Observe the out-going foreign tree edges from $x$
are due to the foreign parents of the incubator $(x, i^*(x))$. 
Each incubator can have at most $k$ foreign parents; hence the out-degree of $x$ due to them is
at most $k$.  



Now we bound the out-degree of a point $x$ due to cross edges.  Recall
that a cross edge between incubators induces
an edge between the identities of the incubators.  Fix some color $c$, and suppose $z$ has color $c$ and
$\{x, z\}$ is a cross edge in $H$
directed from $x$ to $z$; let $i = i^*(x) \leq i^*(z)$.

Observe that $z \in N_i^c$ and a cross edge
that induces $\{x,z\}$ must be at level at most $i$;
hence, $d(x,z) \leq \gamma \cdot r_i$.  Since
$N_i^c$ is an $r_i$-packing,
by Proposition~\ref{prop:small_net} the number of 
such possible $z$'s
is $\gamma^{O(\dim)} = (34 + \frac{272}{\eps})^{O(\dim)}$.

Since there are at most $k + 1$ colors, the number of cross edges coming out of $x$ is bounded by 
$(k + 1) \cdot (34 + \frac{272}{\eps})^{O(\dim)} = \eps^{-O(\dim)} \cdot k$.
\end{proof}

Then, we show that $H$ is a $k$-fault-tolerant $(1 + \eps)$-spanner for $X$. 

\begin{lemma}[Small distance between incubators and occupying zombies] \label{lemma:zombie}
Suppose the incubator $y=(x,i)$ is occupied by
a zombie with identity $z$.  Then, 
the distance $d(x,z)$ is at most $2 r_i$.  In particular,
this implies that if $y'$ is a child of $y$,
then the distance between their occupying zombies
is at most $4 r_i$.
%
\end{lemma}
\begin{proof}
We only need to prove the first
statement about $d(x,z)$. The second
statement follows immediately because $d(y,y') \leq r_i$.

Observe that both $y$ and $z$ have the same color, say color $c$.  We let $x_0 = z$ and $y_0 = (z, 0)$ be the leaf incubator identified by $z$.
For $0 < j \leq i$, let $y_j = (x_j, j)$ be the ancestor of $y_0$ in $T_c$.
(Note that $y_j$ and $y_{j'}$ may refer to the same super incubator for $j \neq j'$.)
By the way that zombies climb, $y_i = y$.
Also, by the construction of incubator trees,
$d(x_j, x_{j-1}) \leq r_j$, for all $0 < j \leq i$.
Hence, $d(x, z) = \sum_{j = 1}^{i}d(x_{j-1}, x_j) \leq \sum_{j=1}^i r_j \leq 2r_i$.
\end{proof}

\begin{figure}[!htb]
\begin{center}
\begin{minipage}[t]{0.35\textwidth}
	\centering
	\includegraphics[width=0.8\textwidth]{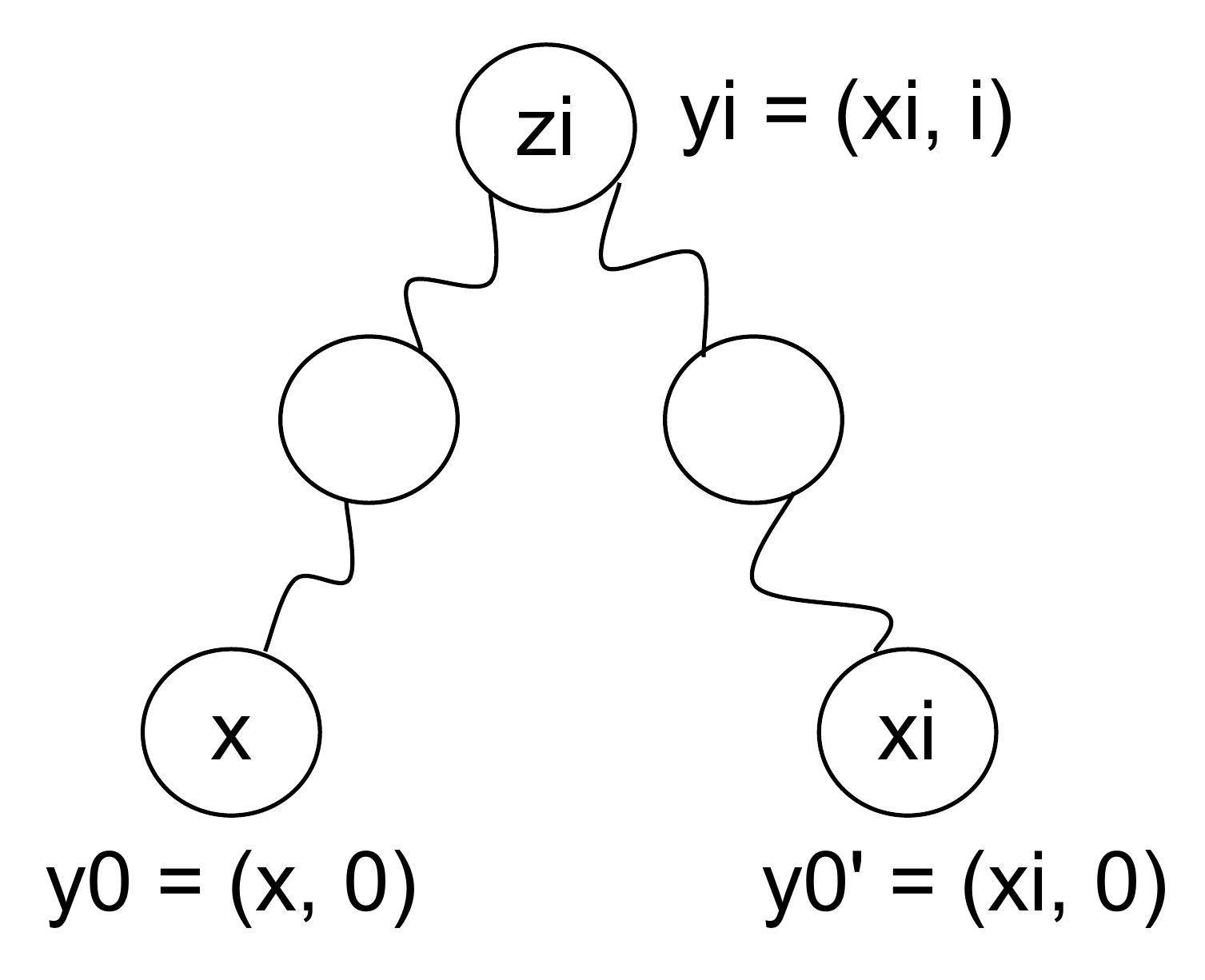}
	\caption{A $c$-\emph{path}}
	\label{fig1}
\end{minipage}
\begin{minipage}[t]{0.4\textwidth}
	\centering
	\includegraphics[width=0.8\textwidth]{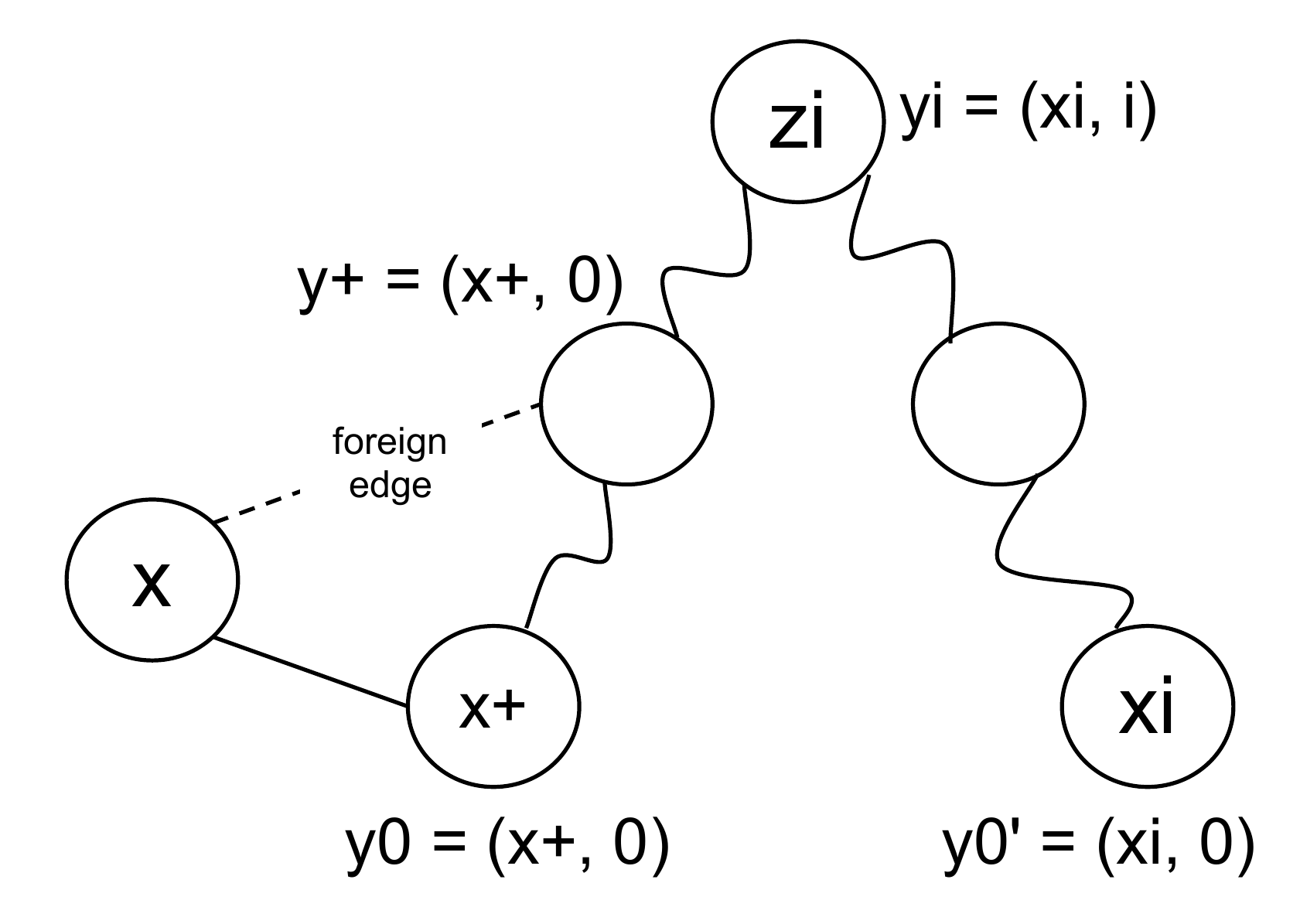}
	\caption{A $c$-\emph{path} following a foreign edge, where $y^+$ stands for 
	$y_{i^*+1}$ and $x^+$ stands for $x_{i^*+1}$.}
	\label{fig2}
\end{minipage}
\end{center}

\end{figure}

\begin{lemma}[Every Level Every Color is Reachable] \label{lemma:cpath}
For any $0 \leq i \leq \ell$ and $x \in X \setminus N_i$,
and for all $c \in [k + 1]$,
there is a path $P_c$ in $H$ between $x$ and the identity of a level-$i$ incubator $y$, with length at most $17r_i$.
In addition, if $x$ is of color $c$, then $P_c$ is a $c$-\emph{path};
otherwise, $P_c$ is the concatenation of a foreign tree edge $\{x,x'\}$
and a $c$-\emph{path}, for some $x'$ with color $c$.
\end{lemma}
\begin{proof}
Suppose $x$ has color $c$.
Let $P^+_c$ be the path in $H$ induced by the path in $T_c$ between the leaf incubator $y_0 = (x, 0)$
and $y_0$'s ancestor at level $i$.
For $0 \leq j \leq i$, denote by $y_j$ the ancestor of $y_0$ in $T_c$ at level $j$,
and let $z_j$ be the zombie occupying $y_j$.
Note that $z_0 = x$.
By Lemma~\ref{lemma:zombie}, for $j \geq 1$, $d(z_{j-1}, z_j) \leq 4r_j$.
Hence, we have $\length(P^+_c) = \sum_{j=1}^{i} d(z_{j-1}, z_j)
\leq \sum_{j=1}^{i} 4 r_j \leq 8 r_i$.

Let $P^-_c$ be the path in $H$ induced by the path in $T_c$ between the incubator $y_i$ and 
its leaf descendant $y'_0$, such that $y'_0$ has the same identity as $y_i$.  
By similar analysis, $\length(P^-_c) \leq 8r_i$. 

Define the path $P_c$ as
the concatenation of $P^+_c$ and $P^-_c$,
which is a $c$-path in $H$ from $x$ to the identity of $y_i$.
We have $\length(P_c) \leq \length(P^+_c) + \length(P^-_c) \leq 16r_i$. 
See Figure~\ref{fig1} for illustration.

Now suppose $x$ has a color $c' \neq c$.
Let $i^*$ be the highest level such that $x \in N_{i^*}$.
Since $x \notin N_i$, we know that $i^* < i$.
Let $y = (x, i^*)$ be an incubator.
By the construction of incubator trees,
there is a foreign parent $y_{i^*+1} = (x_{i^*+1}, i^*+1)$ of $y$ in $T_c$.
Then, we know that $H$ contains an edge $x, \{x_{i^*+1}\}$
induced by the foreign edge between incubators $y$ and $y_{i^*+1}$.

For $i \geq i^*+1$,
let $y_i$ be $y_{i^*+1}$'s ancestor in $T_c$ at level $i$.
Since $x_{i^*+1}$ has color $c$, there is a $c$-path in $H$, denoted by $P'_c$,
which connects $x_{i^*+1}$ and the identity of $y_i$. Furthermore, $P'_c$ has length at most $16r_i$.
Let $P_c$ be the concatenation of $x, \{x_{i^*+1}\}$ and $P'_c$. 
Since $d(x_{i^*+1},x)\leq r_{i^*+1} \leq r_i$,
we know that $\length(P_c) \leq 17r_i$, 
See Figure~\ref{fig2} for illustration.
\end{proof}

\begin{lemma}[Fault-Tolerant Stretch] \label{lemma:climbing}
Let $S \subseteq X$ be a set of at most $k$ failed nodes.
Then, $H \setminus S$ is a $(1 + \eps)$-spanner for $X \setminus S$.
In addition, the $(1 + \eps)$-spanner path in $H \setminus S$
between any $x$ and $y$ contains at most one pure cross edge,
and at most two foreign tree edges.
\end{lemma}
\begin{proof}
Fix $x \neq y \in X \setminus S$.
Let $i$ be such that $r_i < d(x, y) \leq r_{i+1}$,
and let $q := \ceil{\log \frac{68}{\eps}}$.

If $i \leq q - 1$, then $d(x, y) \leq r_{i+1} \leq 2^q \leq \frac{136}{\eps} \leq \gamma \cdot r_0$.
Hence, there is an edge between $x$ and $y$ induced by a cross edge at level $0$.

Suppose $i > q - 1$. Let $j := i - q \geq 0$.
We first show that there is path $P_1$ in $H \setminus S$ with length at most $17r_j$
between $x$ and the identity $x'$ of an incubator at level $j$.
Since there are at most $k$ failed points, i.e., $|S| \leq k$,
there must be a color $c$ such that no failed points has color $c$.
If $x$ is a net point at level $j$,
then there is an incubator $\bar{x} = (x, j)$ at level $j$.
We simply let $x' := x$ and $P_1 := \{x\}$,
and the desired properties hold trivially.
Otherwise, by Lemma~\ref{lemma:cpath},
$x$ is connected to the identity $x'$ of an incubator at level $j$ by
a path $P_1 = \{x, x_1\} \oplus P'_1$ in $H$ with length at most $17r_j$,
where $x_1$ is a point of color $c$
and $P'_1$ is a $c$-path in $H$ between $x_1$ and $x'$.
Note thta all points on $P'_1$ have color $c$ and thus none of them fails.
Hence, it follows that $P_1 \in H \setminus S$.

By similar arguments, we can show that 
there is path $P_2$ in $H \setminus S$ with length at most $17r_j$
between $y$ and $y'$ which is the identity of an incubator at level $j$.

Note that $d(x, x') \leq \length(P_1) \leq 17r_j$
and $d(y, y') \leq \length(P_2) \leq 17r_j$.
It follows that $d(x', y')
\leq d(x', x) + d(x, y) + d(y, y')
\leq r_{i+1} + 34r_j = (2^{q+1} + 34) r_j
< (34 + \frac{272}{\eps}) r_j = \gamma \cdot r_j$.
Hence, there is an edge between $x'$ and $y'$ 
induced by a cross edge between the incubators containing them at level $j$.

Next we show that we can obtain a $(1 + \eps)$-spanner path by
first going from $x$ to $x'$ by $P_1$, then going along the edge $\{x', y'\}$,
and finally going from $y'$ to $y$ by $P_2$:
$d_{H\setminus S}(x, y) \leq \length(P_1) + d(x', y') + \length(P_2)
\leq 17r_j + d(x',x) + d(x,y) + d(y, y') + 17r_j = d(x, y) + 68 r_j
= d(x, y) + \frac{68}{2^q} \cdot r_i \leq (1 + \eps)d(x,y)$.
Note that the $(1 + \eps)$-spanner path contains at most two edges induced by foreign edges,
and the only possible pure cross edge that it
can contain is $\{x',y'\}$.
\end{proof}

\section{Achieving Small Hop-Diameter and Lightness}\label{sec:diam}

Consider the spanner $H$ described in the previous section.
Given a set $S$ of at most $k$ failed nodes,
Lemma~\ref{lemma:basicH} states that for any points $x, y\in X\setminus S$, 
there exists a $(1+\eps)$-spanner path in $H\setminus S$ between $x$ and $y$
which is the concatenation of at most two foreign edges, at most one pure cross 
edge, and at most two $c$-\emph{paths} for some $c$. 
Denote this path as $P_{H\setminus S}(x,y)$.
Notice that a $c$-\emph{path}
is induced by some path in the incubator tree $T_c$. 
In this section, we add edges to shortcut the incubator trees,
such that each of the above $c$-paths can be
replaced with a subpath with at most $O(\log n)$ hops.

In \cite{DBLP:conf/esa/SolomonE10},
a construction for $1$-spanners for tree metrics with short hop-diameter is introduced.
.
\begin{fact}[Spanner Shortcut \cite{DBLP:conf/esa/SolomonE10}]\label{fact:short}
Let $T$ be a tree (whose edges have non-negative weights) with $n$ nodes and maximum degree $D$. 
For the tree metric induced by the shortest-path distances in $T$, there exists a $1$-spanner 
$J$ with $O(n)$ edges, maximum degree at most $D + 4$,
hop-diameter $O(\log n)$ and lightness $O(\log n)$. 
Furthermore, $J$ has the following properties.
\begin{compactitem}
\item[1.]
Spanner $J$ includes all edges in the tree, i.e., $T\subseteq J$.
\item[2.]
The construction of $J$ only depends on the structure of $T$, and is independent
of the edge weights.
\item[3.]
Any path in $T$ can be replaced by a subpath consisting of at most $O(\log n)$ edges in $J$.
\end{compactitem}
\end{fact}

We also have the following lemma.
\begin{lemma}[Tree edges and cross edges are light]\label{lemma:lightness} 
Let $E_i$ be the set of tree edges between level $i$ incubators 
and level $i+1$ incubators, 
and $C_i$ be the set of cross edges between level $i$ incubators.
By convention, let $E_\ell := \emptyset$. 
Let $\widehat{r} = \frac{k^2 \Delta}{n^2\gamma}$, and define $\sigma := \floor{\log_2 \widehat{r}}$.  Then,

\begin{compactitem}
\item[1.]
for any $\sigma < i\leq \ell$, the total weight of the edges induced by $E_i\cup C_i$ 
is $\eps^{-O(\dim)}\cdot k^2 \cdot w(\MST)$;
\item[2.]
the total weight of the edges induced by $\bigcup_{i=0}^{\sigma} (E_i\cup C_i)$ is $O(k^2 \cdot w(\MST))$. 
\end{compactitem}
\end{lemma}
\begin{proof}
First, we analyze the total weight of edges induced by $E_i$, for any $\sigma < i \leq \ell$. 
Since $N^c_i$ is $r_i$-packing for any $c\in [k+1]$ from Fact~\ref{fact}, 
it holds that $w(\MST) \geq \frac{1}{2} \cdot r_i\cdot |N^c_i|$. 
Consequently, $w(\MST) \geq \frac{1}{2k} \cdot r_i\cdot |N_i|$. 
Notice that the total weight of edges induced by $E_i$ is at most $2(k+1) \cdot r_i\cdot |N_i|$, which
is $O(k^2 \cdot w(\MST))$. 

Consider the edges in $E_i$ with $0 \leq i\leq \sigma$. 
For each subtree rooted at level-$\sigma$ incubator in $T_c$, 
the total weight of edges induced by it is bounded by
\[n\cdot r_{\sigma} \sum_{i=0}^{\ell} \left(\frac{1}{2}\right)^i \leq \frac{2 k^2\cdot \Delta}{n\cdot \gamma}.\]
Since there are at most $n$ such subtrees, we know that the total weight of edges induced by 
$\bigcup_{i=0}^{\sigma} E_i$ is $O(k^2 \cdot w(\MST))$. 

Then we analyze the cross edges.
Consider an incubator $y$ at level $i$.
The number of cross edges
incident to $y$ in $C_i$ at most is $(k+1) \cdot \gamma^{2\dim}$, each of which has weight at most 
$\gamma\cdot r_i$. Hence, the total weight of edges induced by $C_i$ is 
$\gamma^{2\dim+1}\cdot (k+1)\cdot r_i\cdot |N_i|$ which is $\eps^{-O(\dim)} \cdot k^2 \cdot w(\MST)$. 

In particular, for $i = \sigma - j$ with $0\leq j \leq \sigma$,
there are at most $n^2$ cross edges at level $i$,
each of which has weight at most $\gamma\cdot \left(\frac{1}{2}\right)^j\cdot \widehat{r}$.
Hence, the total weight of edges induced by $\bigcup_{i=0}^{\sigma} C_i$ is at most 
\[\sum_{j=0}^\sigma n^2\cdot \gamma\cdot \widehat{r}\cdot \left(\frac{1}{2}\right)^j\leq 2 k^2 \cdot \Delta 
= O(k^2 \cdot w(\MST)).\]
\end{proof}
\noindent \textbf{Shortcutting Sub-Trees at Level below $\sigma$.} In Lemma~\ref{lemma:lightness}, we have $l - \sigma = O(\log n)$, 
and the total weight of induced edges below level $\sigma$ is $O(k^2 \cdot w(\MST))$. 
We shortcut each subtree rooted at each level $\sigma$ incubator, with a spanner given by Fact~\ref{fact:short}. 
Notice that this procedure adds edges to shortcut the paths consisting of local trees edges. 
Hence, for any $S\subseteq X$ of size at most $k$, and points $x, y\in X$, 
the $c$-path in $P_{H\setminus S}(x,y)$ can be substituted by its subpath with at
most $O(\log n)$ hops. Furthermore, the lightness of our spanner over $X$ is increased
by a factor of at most $O(\log n)$.

\noindent\textbf{Skeleton edges.} The edges between incubators added by Fact~\ref{fact:short},
together with the local tree edges are called \emph{skeleton edges}. Recall our naming system implies that an
edge between points in $X$ is a skeleton edge, if it is induced by a skeleton edge between incubators.

Define a new spanner $\widehat{H}$ to be the union of $H$ and the skeleton edges. 
Combing Fact~\ref{fact:short} and Lemma~\ref{lemma:lightness}, we have the following
lemma, which is the main result of this section.

\begin{lemma}\label{lemma:hopH}
The spanner $\widehat{H}$ is a $(k, 1 + \eps)$-VFTS for $X$ with the following properties.
\begin{compactitem}
\item[1.]
For any set $S\subseteq X$ with $|S| \leq k$, and any $x, y \in X\setminus S$, 
there exists a $(1 + \eps)$-spanner path in $\widehat{H} \setminus S$ between $x$ and $y$
which is the concatenation of at most two foreign tree edges, at most one pure cross edge, 
and at most two paths with $O(\log n)$ skeleton edges.
\item[2.]
The maximum degree of $\widehat{H}$ due to skeleton edges is bounded by $2^{O(\dim)}$.
\end{compactitem}
\end{lemma}

\section{Achieving Small Degree via Single-Sink Spanners}
\label{sec:degree}

Up to this point,
the spanner we have constructed has $(1 + \eps)$-stretch,
hop-diameter $O(\log n)$ and lightness $\eps^{-O(\dim)}\cdot k^2\cdot \log n$,
but may suffer from a large maximum degree.
In the next two sections,
we reduce the maximum degree in the spanner to a constant,
with the sacrifice of increasing a factor of $k$ in the lightness.

Our technique of reducing degrees in fault-tolerant spanners is based on single-sink spanners.
Given a root point $v \in X$, a spanner $H$ for $X$ is a \emph{$k$-vertex-fault-tolerant $v$-single-sink $t$-spanner} ($(k, t, v)$-VFTssS), 
if for any subset $S \subseteq X \setminus \{v\}$ with $|S|\leq k$, 
and any point $x \in X \setminus S$, 
it holds that $d_{H \setminus S}(v, x) \leq t \cdot d(v, x)$,
where $t$ is called the \emph{root-stretch} of $H$.
In~\cite{Chan2012}, a construction of a $(k, 1 + \Theta(\eps), v)$-VFTssS with maximum
degree $O(k)$ for doubling metrics is given, but there is no guarantee on hop-diameter.
We review the main ideas of the previous fault-tolerant single-sink spanner 
construction~\cite{Chan2012} and show how it
can be modified to achieve both $O(\log n)$ hop-diameter and small weight,
as stated in Theorem~\ref{th:single_sink}.

\begin{theorem}[Fault-Tolerant Single-Sink Spanner]
\label{th:single_sink}
Given $0 < \epsilon \leq \frac{1}{6}$ and a sink $v \in X$,
there exists a $(k, 1 + 10 \epsilon, v)$-VFTssS $H_v$ with maximum degree $(\frac{1}{\epsilon})^{O(\dim)} k$,
hop-diameter $O(\log n)$ (with respect to sink $v$) and weight at most $(1+\epsilon)(k+1) \sum_{x \in X} d(x,v)$.
\end{theorem}
The first idea (also used in~\cite{DBLP:conf/soda/ChanGMZ05}) is that instead of 
connecting every point in $X$ directly to the sink $v$,
a point far away from $v$ can be connected to $v$ with a path in which
distances of points from the sink decrease geometrically rapidly.

\begin{lemma}[Path with Rapidly Geometrically Decreasing Distances]
\label{lemma:rapid_geom}
Suppose $0 < \epsilon \leq \frac{1}{3}$, and $P$ is a path $v=p_0, p_1, \ldots, p_l=x$
such that for all $i \geq 1$, $d(v,p_i) \leq \epsilon d(v, p_{i+1})$.  Then,
the path distance $d_P(v,x) \leq (1 + 3 \epsilon) d(v,x)$.
\end{lemma}

\begin{proof}
The result follows from a simple induction proof on $l$.  For $l=1$, the result is trivial.  Assume that
$d_P(v,p_l) \leq (1 + 3 \epsilon) d(v, p_l)$.  By the hypothesis of the lemma,
we have $d(v,p_l) \leq \epsilon d(v, p_{l+1})$ and so
by the triangle inequality,
 $d(p_{l+1}, p_l) \leq d(p_{l+1},v) + d(v,p_l) \leq (1 + \epsilon) d(v,p_{l+1})$.  
 
 Therefore, $d_P(v, p_{l+1}) \leq d_P(v, p_l) + d(p_l, p_{l+1}) \leq (1 + 3 \epsilon) \epsilon \cdot d(v, p_{l+1})
 + (1 + \epsilon) d(v,p_{l+1}) \leq (1 + 3 \epsilon) d(v,p_{l+1})$,
 where the last inequality holds because $\epsilon \leq \frac{1}{3}$.
\end{proof}

Since we consider points whose distances from the sink decrease
geometrically rapidly, \emph{ring partition} is considered in~\cite{Chan2012}.

\noindent \textbf{Ring Partition.}
Let $\ell = \ceil{\log_\frac{1}{\eps}\Delta}$ and $r_i = \frac{1}{\eps^i}$ with $i\in [\ell]$.
For convenience, let $r_0 = 1$ (recall that we assume inter-point distances are larger than 1).
Consider the rings, denoted by $R_1, \ldots, R_{\ell}$, where $R_i:=R(v, r_{i-1}, r_i)$.
For convenience, let $R_0 := \{v\}$.  We say a point is in ring $i$ if it is in $R_i$.
The rings are pairwise disjoint and their union covers $X$.  We say that a point $x$ is
\emph{at least 2 rings below} another point $y$, if for some $i \leq j-2$, $x \in R_i$ and $y \in R_j$;
in this case, observe that $d(v,x) \leq \epsilon \cdot d(v,y)$.
In view of Lemma~\ref{lemma:rapid_geom}, we would like to connect a point to the sink
through a path such that the next point is at least 2 rings below the previous point.  

However,
the degree of the resulting spanner could be large if there is some ring that contains a lot of points.
Hence, in~\cite{Chan2012}, an appropriate net is constructed for each ring and each ring is decomposed into clusters, each of
which is covered by the closest net point. In particular, for each $i\in [\ell]$, we build an $\eps r_{i-1}$-net $N_i$ for $R_i$.
By Proposition~\ref{prop:small_net}, $N_i$ contains at most $(\frac{r_i}{\eps r_{i-1}})^{2\dim} = \eps^{-4\dim}$ points.
We denote this upper bound by $\Gamma:= \ceil{\eps^{-4\dim}}$ and then we have $|N_i| \leq \Gamma$.
Let $N := \cup_{i > 0}N_i$ be the set of net points.
Then, for each net point $y\in N_{i}$, we construct a \emph{net cluster} $C_y$, such that a point $x \in X$ is in $C_y$ \emph{iff} $x$ is in $R_i$,
and among all points in $N_i$, $y$ is the closest one to $x$ (breaking ties arbitrarily).

\noindent \textbf{Achieving Fault-Tolerance via Multiple Portals for Each Cluster.}  If all nodes do not fail,
then we can first build a single-sink spanner for the net points in all rings, and then
for each cluster, we can build a subgraph such that each point is connected to the net point with a short path.
However, the net point might fail, and 
so for each $y \in N$, we arbitrarily choose $k + 1$ \emph{portals} $Q_y \subseteq C_y$ (if $|C_y| < k + 1$, we let $Q_y = C_y$); we can think of in addition to the net point $y$, we choose $k$ extra points (if there are enough points) in the cluster
as portals such that when $k$ nodes fail, there will be at least one functioning portal to connect that cluster with the outside world.
Let $Q_i := \cup_{y \in N_i}Q_y$ be the portals in $R_i$, and let $Q := \cup_{y \in N}Q_y$ be the set of all portals.
Note that $|Q_i| \leq (k + 1) \cdot |N_i| \leq \Gamma \cdot (k + 1)$.

The following approach is used in~\cite{Chan2012} to build
a single-sink spanner for $X$ by combining single-sink spanner for the portals
and subgraphs connecting points in each cluster to their portals.

\begin{lemma}[Single-Sink Spanner for Portals and Connecting Points to Portals]
\label{lemma:single_sink_portal_connect}
Let $0 < \epsilon \leq \frac{1}{6}$.
Suppose $H_Q$ is a $(k,1+3\epsilon, v)$-VFTssS for $Q$ with maximum degree $\delta_1$ and hop-diameter $D_1$. 
Moreover, for each $y \in N$ such that the cluster $C_y$ has radius $r$,
suppose $H_y$ is a subgraph on $C_y$ with maximum degree $\delta_2$
such that for any subset $S$ of vertices of size $k$, for any $x \in C_y \setminus S$,
there exists a portal $q \in Q_y \setminus S$ such that there is a path
from $x$ to $y$ with length $4r$ and $D_2$ hops.
Then, the subgraph $H_v := H_Q \cup (\cup_{y \in N} H_y)$ is a $(k, 1 + 10 \epsilon, v)$-VFTssS for $X$
with maximum degree $\delta_1 + \delta_2$ and hop-diameter $D_1 + D_2$.
\end{lemma}

\begin{proof}Suppose $S$ is a set of $k$ failed nodes and $x \notin S$.
We can assume that $x \neq v$; otherwise the conclusion holds trivially.  Suppose $x \in R_i$ (where $i \geq 1$), and
let $y \in N_i$ be the net point covering $x$, i.e., $x \in C_y$.  Hence, the
radius of $C_y$ is $r = \epsilon r_{i-1} \leq \epsilon d(v,x)$.

From assumption, there exists a path $P_2$ in $H_y \setminus S$ from $x$ to some
functioning portal $q \in Q_y$ with $D_2$ hops and length $4r \leq 4 \epsilon \cdot d(v,x)$.
Note that $d(v,q) \leq d(x,v) + d(x,q) \leq (1 + 2 \epsilon) d(v,x)$.

By assumption, since $H_Q$ is a fault-tolerant spanner for $Q$,
there exists a path $P_1$ in $H_Q \setminus S$ from $q$ to $v$
with $D_1$ hops and length $(1 + 3 \epsilon) d(q,v) \leq (1 + 6 \epsilon) d(v,x)$,
where the last inequality holds because $\epsilon \leq \frac{1}{6}$.

Hence, it follows that $P_2$ and $P_1$ forms a path from $x$ to $v$
that has length at most $(1 + 10 \epsilon) d(x,v)$.  The maximum degree of
a point in $H$ is trivially at most $\delta_1 + \delta_2$, since the clusters $C_y$'s
contain disjoint points.
\end{proof}

\subsection{Fault-Tolerant Single-Sink Spanner for Portals}

We review the fault-tolerant single-sink spanner construction~\cite{Chan2012} for
portals and describe how we can achieve $O(\log n)$ diameter.  Recall that
because of Lemma~\ref{lemma:rapid_geom}, when we traverse on a path from some point $x$ to $v$,
we would like the next point to be at least 2 rings below that of the previous point.  At the
same time, we would like to achieve fault-tolerance.  Hence, we arrange the portals in groups of size $k+1$
sorted in non-decreasing distance from $v$.

\noindent \textbf{Grouping Portals.}
We sort points in $Q = \{q_1, q_2, \ldots, q_m\}$ in non-decreasing distance from $v$, i.e., $d(v, q_1) \leq d(v, q_2) \leq \cdots \leq d(v, q_m)$, where $m := |Q|$.
For convenience, let $q_0 := v$. Then, we divide the points in $Q$ into groups of size $k+1$. 
Specifically, let $A_j := \{q_l \in Q | (j - 1) \cdot (k + 1) + 1 \leq l \leq j \cdot (k + 1)\}$ for $j \geq 1$.
For convenience, let $A_0 := \{v\}$, and for $i < 0$, $A_i := A_0$.

Observe that any consecutive two rings can contain at most $2 \Gamma (k+1)$ portals.  Hence
for any $1 \leq i \leq j-2\Gamma - 1$, any point in $A_i$ must be at least two rings below any point in $A_j$.
Hence, for each $j \geq 1$, we can connect group $A_j$ with $A_{j - 2 \Gamma -1}$,
where connecting two groups $A_i$ and $A_j$ means adding edges between every point in $A_i$ and every point in $A_j$.
Observe that for $i < j$, when $k$ nodes fail, there is at least one node in $A_i$ available for the nodes in $A_j$ to reach $v$.
Note that the degree of $v$ is at most $(2 \Gamma + 1)(k+1)$, and every other node has degree at most $2(k+1)$.
This is the construction given in~\cite{Chan2012} and according to Lemma~\ref{lemma:rapid_geom},
the stretch to $v$ is at most $1 + 3 \epsilon$.  However, the hop-diameter could be large,
because for every $i \geq 1$, every group $A_i$ is connected to at most one group $A_j$ with $j > i$.

We modify our construction in the following way: for every $j \geq 1$,
we connect $A_j$ to $A_{\ceil{\frac{j- 2 \Gamma - 1}{2}}}$.  As before, when
we connect two groups, we add edges to form a complete bipartite graph between them.  Suppose $H_Q$
is the resulting subgraph on the portals $Q$.

\begin{lemma}[Fault-Tolerant Single-Sink Spanner for Portals]
\label{lemma:single_sink_portal}
For $0 < \epsilon \leq \frac{1}{3}$, the subgraph $H_Q$ is a $(k, 1 + 3\epsilon, v)$-VFTssS
for $Q$ with maximum degree $O(\Gamma k)$ and hop-diameter $O(\log n)$.
\end{lemma}

\begin{proof}
The property of $k$-fault-tolerance follows from the previous construction.  The stretch analysis is the same as before
because for each $j \geq 1$, group $A_j$ is connected to $A_i$, where $i \leq j - 2 \Gamma - 1$.
Moreover, the degree of $v$ is still $O(\Gamma k)$.

If we view $A_0$ is the root, and view $A_i$ as $A_j$'s parent if $A_j$ is connected to $A_i$ with $i < j$,
then we can see that every group apart from the root has at most two children.  Hence, the degree of a node
other than $v$ is at most $3(k+1)$.

Moreover, observe that for $j \geq 1$, for all but at most one internal group $A_j$, the group $A_j$ is connected
to two child groups.  Hence, it follows that it takes $O(\log n)$ hops to reach the root $A_0$ from any group.
\end{proof}

\subsection{Connecting Points in Each Cluster to Portals}

For a net point $y \in N$, we describe
how to build a subgraph $H_y$ on the cluster $C_y$
to connect points in $C_y$ to the portals $Q_y$ such
that the conditions stated in Lemma~\ref{lemma:single_sink_portal_connect}
are satisfied.

\noindent \textbf{Connecting Points in Clusters to Portals.}
Recall that the cluster $C_y$
has radius $r$ centered at $y$.
As in~\cite{Chan2012}, we define a procedure $\Add(C_y, Q_y, r)$ which adds edges to connect points in $C_y$ with portals in $Q_y$,
where $C_y$ is a cluster with radius $r$ centered at some point.

The idea in~\cite{Chan2012} is that we partition the points in $C_y \setminus Q_y$
into sub-clusters with radius $\frac{r}{2}$ and recurse on the sub-clusters.  For each sub-cluster
$C_z$, we choose $k+1$ points $Q_z$ (if there are enough points) in $C_z$ as portals; we connect every point in $Q_y$ 
with every point in $Q_z$, and recursively call $\Add(C_z, Q_z, \frac{r}{2})$.  The
base case of the recursion is when a cluster has no more points other than its portals, i.e., $C_z = Q_z$.

However, it is possible that the points are distributed such that at every level of recursion, there
is only one sub-cluster in the next level; this means the number
of levels of recursion can be linear in the size of the original cluster.
 Notice that the number of hops for a portal at the last level to reach a portal at the first level
 is equal to the number of levels of recursion.  A simple modification is to first divide the points in $C_y \setminus Q_y$
evenly into two sets $U_1$ and $U_2$, i.e., $||U_1| - |U_2|| \leq 1$; then we can
work on $U_1$ and $U_2$ as described above.  The formal procedure is described below.


 
\SetAlgorithmName{Procedure}{Name}{Procedure}
\LinesNumbered
\begin{algorithm}[h]
\SetAlgoLined
\KwResult{Adds edges to connect points in $C_y$ with portals in $Q_y$}
If $C_y = Q_y$, \textbf{return}\; \tcp{Now suppose $Q_y \subsetneq C_y$}
Partition the points in $C_y \setminus Q_y$ evenly into two sets $U_1$ and $U_2$ such that $||U_1| - |U_2|| \leq 1$\;
\ForEach{$i \in \{1, 2\}$}{
	Build an $\frac{r}{2}$-net $N_i$ for $U_i$\;
	\ForEach{$z \in N_i$}{
		Build a sub-cluster $C_z := \{x \in U_i : z$  is the closest point in $N_i$ to  $x\}$\;
		Arbitrary select $k + 1$ portals $Q_z$ in $C_z$ (select all points in $C_z$ if $|C_z| < k + 1)$\;
		Add an edge between every point in $Q_y$ and every point in $Q_z$\;
		Recursively call $\Add(C_z, Q_z, \frac{r}{2})$\;
	}
}
\caption{$\Add(C_y, Q_y, r)$}
\label{fig:clusters}
\end{algorithm}

%

Let $H_y$ be the resulting subgraph returned by $\Add(C_y, Q_y, r)$. We have the following lemma. 

\begin{lemma} \label{lemma:cluster}
The graph $H_y$ has maximum degree $2^{O(\dim)} \cdot k$.
Suppose $S \subseteq X$ be a set of at most $k$ points. For any $x \in C_y \setminus S$ and any $q \in Q_y \setminus S$,
there exists a path $P_2$ in $H_y \setminus S$ between $q$ and $x$ with $\length(P_2) \leq 4r$ and $O(\log n)$ hops.
\end{lemma}

\begin{proof}
For each $i \in \{1, 2\}$,
$N_i$ is a $\frac{r}{2}$-net for $U_i$,
and is contained in the ball centered at $y$ of radius $r$.
By Proposition~\ref{prop:small_net},
$|N_i| \leq 2^{2\dim}$.
Hence, each point in $Q_y$ is connected to portals
in at most $2^{2\dim + 1}$ sub-clusters.
Since each sub-cluster has at most $k + 1$ portals,
it follows that each point in $Q_y$ is connected to
at most $2^{2\dim + 1}(k + 1)$ portals in sub-clusters.
In addition, points in $Q_y$ may also be connected to $k + 1$ portals in $C_y$'s super-cluster.
Hence, each point in $Q_y$ is connected to at most $(2^{2\dim + 1} + 1)(k+1) = 2^{O(\dim)}\cdot k$ points in $H_y$.

The $k$-fault-tolerance properties follows because if $C_z$ is a sub-cluster of $C_y$,
then even when $k$ nodes fail, there always exists a functioning portal in $Q_y$ connecting
to the functioning portals in $Q_z$.

Observe that the distance between a portal in the first level and one
at the next level is at most $2r$, and this decreases geometrically with factor $\frac{1}{2}$.  Hence,
a path $P_2$ from a a portal in the first level to a portal in the last level has length at most $4r$.  Moreover,
because we divide the remaining points evenly at each level of recursion,
the total number of levels of recursion is $O(\log n)$, and so the path $P_2$ has $O(\log n)$ hops.
\end{proof}

\subsection{Weight Analysis}
\begin{lemma} \label{lemma:weight}
The spanner $H_v := H_Q \cup (\cup_{y \in N} H_y)$ constructed above has weight at most 

$(1 + \epsilon)(k + 1)\sum_{x \in X}d(x, v)$.
\end{lemma}

\begin{proof}
For each edge in $H_v$, we charge the edge to one of its endpoints.  Then, we show that
for each $x \in X$,
the sum of the weights of the edges charged to $x$ is at most $(1+\epsilon)(k+1) d(x,v)$.

For an edge in $H_Q$, we charge it to the portal in the higher ring. For an 
edge in some $H_y$, it connects a portal $u$ in some cluster $C$ to a portal $v$
in one of its sub-clusters; we charge this edge to $v$.  Observe that for all $x \neq v$,
at most $k+1$ edges are charged to $x$ and they are either 
 all of the first type or all of the second type.
 
Consider a node $x$ for which all edges charged to it are in $H_Q$.  It follows
that for each such edge $\{x,y\}$, the point $y$ must be at least 2 rings below $x$.
Hence, $d(y,v) \leq \epsilon d(x,v)$, and so $d(x,y) \leq d(x,v) + d(v,y) \leq (1+\epsilon) d(x,v)$.

Consider a node $x$ for which all edges charged to it are in $H_y$ for some $y \in N$.
Suppose $x$ is in ring $i$.  Then, it follows that $x$ in some cluster $C_y$ with radius
$r \leq \epsilon d(x,v)$.  Hence, every edge charged to $x$ has
weight at most $2r \leq 2 \epsilon d(x,v) \leq (1 + \epsilon) d(x,v)$.

Since for every $x \neq v$, at most $k+1$ edges are charged to it,
it follows the sum of weights of edges charged to $x$ is at most $(1+\epsilon)(k+1) d(x,v)$, as required.
\end{proof}

\section{$(k, 1 + \eps)$-VFTS with Bounded Degree, Hop-Diameter and Lightness}

In this section, we show how to combine our previous construction up to Section~\ref{sec:diam}
and fault-tolerant single-sink spanners in Thoerem~\ref{th:single_sink} to get a fault-tolerant $(1 + \eps)$-spanner
with simultaneously bounded degree, short hop-diameter and small lightness.

We first use the construction up to Section~\ref{sec:diam} 
to build a $(k, 1 + \frac{\eps}{3})$-VFTS $H_0$.
Let $\widehat{E}$ be the skeleton edges,
and $\overline{E}$ be the non-skeleton edges in $H_0$.
Then we direct the edges in $\overline{E}$ using the techniques in 
Section~\ref{sec:zombie} so that
the out-degree of each point $x \in X$ due to $\overline{E}$ is $\eps^{-O(\dim)} \cdot k$.
We denote an edge $\{x, y\}$ by $(x, y)$ if it is directed from $x$ to $y$.
For each $x \in X$, let $N_{in}(x) := \{y \in X : (y, x) \in \overline{E}\}$,
and we build a $(k, 1 + \frac{\eps}{3}, x)$-VFTssS $H_x$ for $N_{in}(x) \cup \{x\}$.
To get the final spanner,
we replace the star consisting of edges in $\overline{E}$ directed into $x$ with $H_x$.
In other words, we let $E := \widehat{E} \cup (\cup_{x \in X}E(H_x))$
and take the graph $H^*$ consisting of all edges in $E$, as the final spanner for $X$.

\begin{theorem}
For $0 < \eps < \frac{1}{2}$, $H^*$ is a $(k, 1 + \eps)$-VFTS.
In addition, its maximum degree is $\eps^{-O(\dim)} \cdot k^2$,
its hop-diameter is $O(\log n)$,
and its lightness is $O(k^3 \log n)$.
\end{theorem}
\begin{proof}
We first show the $1 + \eps$ stretch.
Let $S \subseteq X$ be a set with at most $k$ points.
Since $H_0$ is a $(k, 1 + \frac{\eps}{3})$-VFTS,
for any $x, y \in X \setminus S$,
there is a $(1 + \frac{\eps}{3})$-spanner path
$P_0 \subseteq H_0 \setminus S$ between $x$ and $y$.
Also, by Lemma~\ref{lemma:hopH},
there exists a path $P_0 \in H_0 \setminus S$ 
that contains at most one pure cross edge
and at most two foreign tree edges.
In other words, $P_0$ contains at most $O(\log n)$ edges,
and at most $3$ edges in $\overline{E}$.

Consider an edge $\{u, v\} \in P_0 \cap \overline{E}$, and suppose it is directed as $(u, v)$.
Since $H_v$ is a $(k, 1 + \frac{\eps}{3}, v)$-VFTssS and $u, v$ are both functioning,
there is a $(1 + \frac{\eps}{3})$-spanner path $P_{uv}$ in $H_v \setminus S$ between $x$ and $y$.
Let $P$ be the path obtained by replacing each $\{u, v\} \in P_0 \cap \overline{E}$ with $P_{uv}$.
Then, we know that $P$ is contained in $H \setminus S$
and is a spanner path between $x$ and $y$ with stretch at most $(1 + \frac{\eps}{3})^2 \leq 1 + \eps$.

Next we bound the hop-diameter.
Note that the $(1 + \frac{\eps}{3})$-spanner path $P_0$ between $x$ and $y$ in $H_0$
contains at most $O(\log n)$ edges.
To obtain $P$, at most $3$ edges in $P_0 \cap \overline{E}$ are replaced by
a $(1 + \frac{\eps}{3})$-spanner path in a single-sink spanner.
Recall that the single-sink spanners constructed in Section~\ref{sec:degree} has
hop-diameter $O(\log n)$.
Hence, the replacement only increases the number of edges by $O(\log n)$,
and thus $P$ contains at most $O(\log n)$ edges.
It follows that the hop-diameter of $H^*$ is at most $O(\log n)$.

Now we bound the degree of an arbitrary point $x \in X$.
We already know that the degree of $X$ due to edges in $\widehat{E}$ is $2^{O(\dim)}$,
and we only need to give an upper bound for $x$'s degree due to edges in the single-sink spanners.

The edges in the single-sink spanners incident to $x$ are contained in $H_x$
and $H_y$'s such that there is an edge $(x, y) \in \overline{E}$.
Note that the number of $H_y$'s involving $x$ is bounded by
the out-degree of $x$ due to $\overline{E}$, which is $\eps^{-O(\dim)} \cdot k$.
Also recall that the degree of $x$ in $H_x$ is $\eps^{-O(\dim)} \cdot k$ and the degree of $x$ in each $H_y$ is $2^{O(\dim)} \cdot k$.
It follows that the degree of $x$ due to single-sink spanners is $\eps^{-O(\dim)} \cdot k^2$.
Hence, the degree of $x$ in $H^*$ is $\eps^{-O(\dim)} \cdot k^2$.

Finally, we analyze the lightness.
By Lemma~\ref{lemma:hopH},
$H_0$ has lightness $O(k^2 \log n)$.
Note that the star spanning $N_{in}(x)$ centered at $x$ has weight
$\sum_{y \in N_{in}(x)} d(x, y)$.
By Lemma~\ref{lemma:weight}, the single-sink spanner $H_x$ has weight at most
$O(k) \cdot \sum_{y \in N_{in}(x)} d(x, y)$.
Hence, the replacement of the stars with single-sink spanners increases the weight
by a factor of $O(k)$.
Also observe that the stars replaced are disjoint.
It follows that the lightness of $H^*$ is at most $O(k^3 \log n)$.
\end{proof}

{\footnotesize
\bibliography{spanner}
\bibliographystyle{abbrv}
}



\end{document}